# 通用平台高性能可扩展网络地址转换系统

李峻峰, 李 丹, 黄昱恺, 程 阳, 令瑞林

( 清华大学 计算机科学与技术系, 北京 100084)

**摘要**: 为了提高通用平台网络地址转换(NAT)的性能，设计了高性能可扩展网络地址转换系统 Quick NAT．高速网络地址转换查表算法将原始规则表划分为若干子表，并采用哈希查表算法极大地提高了 NAT 规则查找效率；为了充分发挥多核中央处理器(CPU)与多队列网卡的性能优势，设计高效并行架构，使用本地化的连接记录表和基于比较并交换原子操作的无锁 NAT 规则表，避免多 CPU 核访问修改全局表带来的锁开销；借助轮询取代中断、越过内核等机制，并全程使用指针操作数据包实现零拷贝，进一步降低开销．实验结果表明，Quick NAT 可以极大地提高 NAT 查表的效率和吞吐量，具有较强的多核可扩展性，能够在 10 Gbit/s 的网络环境下实现 64 byte 小包线速．

**关 键 词**: 网络地址转换; 性能优化; 哈希算法; 数据结构本地化

**中图分类号**: TP393　　　　　**文献标识码**: A

## High Performance and Scalable NAT System on Commodity Platforms

LI Jun-feng, LI Dan, HUANG Yu-kai, CHENG Yang, LING Rui-lin

(1. Department of Computer Science and Technology, Tsinghua University, Beijing 100084, China)

**Abstract:** Quick network address translation (NAT) is proposed to improve the network performance of the NAT system on the commodity server by three ways. First, the quick NAT search algorithm is designed to use the Hash search instead of the sequential search to reduce latency when looking up the NAT rule table. Second, to leverage the power of the multi-core central processing unit (CPU) and the multi-queue network interface card, Quick NAT enables multiple CPU cores to process in parallel. The localized connection tracking table and the compare-and-swap based lock-free NAT Hash tables are designed to eliminate the lock overhead. Third, Quick NAT uses the polling and zero-copy delivery to reduce the cost of interrupt and packet copies. The evaluation results show that Quick NAT obtains high scalability and line-rate throughput on the commodity server.

　　随着互联网规模不断增长，越来越多的设备和用户接入网络．截至 2019 年底，全球可供域名系统(DNS, domain name system)查询的互联网域名数目已经超过 10 亿个，网民数目已经超过 41 亿人，超过全世界总人口数的一半[1-2]．然而 IPv4 地址空间早在 2011 年就被全部耗尽[3]，已无法满足实际需求．IPv6 具有更大的地址空间，可以解决 IPv4 地址不够用的问题．但是，由 IPv6 完全取代 IPv4 不仅需要网络基础设施的更新换代，而且需要网络应用的修改升级以适配 IPv6，这些都是广泛部署 IPv6 的巨大障碍[4]．因此，在未来的一段时间，IPv4 仍然会占据主导地位[5]．在 IPv4 网络中，网络地址转换(NAT, network address translation)可以使不同的内网设备共享使用同一个外网 IP 地址，不仅

完美地解决了 IP 地址不足的问题，而且还能够有效隐藏并保护网络内部的计算机，因此被广泛应用于各种类型 Internet 接入方式和各种类型的网络中．NAT 的实现部署方式主要分为专用硬件和软件 2 类．专用 NAT 硬件设备虽然性能高，但是价格昂贵且不便于统一管理．为了方便管理和降低成本，可以在通用平台上借助 Linux 内核的 Netfilter/IPTables 组件实现 NAT．由于 NAT 需要对匹配规则的每一个数据包的源地址或者目的地址进行修改，NAT 的性能影响数据流的完成时间，进而影响应用性能．但是经过实验发现，在网络流量较大时，Linux 内核的 Netfilter/IPTables 性能无法满足需求，成为制约应用性能的瓶颈．

## 1 内核 Netfilter/IPTables 框架

Linux 内核对 NAT 功能的支持是通过 Netfilter/IPTables 框架实现的，Netfilter 是 1 个通用数据包过滤架构，如图 1 所示．

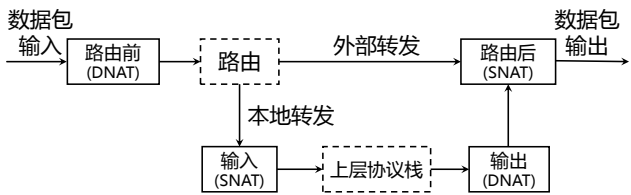

图 1 Netfilter 框架

Netfilter 框架主要由 5 个 Hook 点构成[6]，各个 Hook 点可以挂载回调函数对数据包进行处理[7]，并根据回调函数的处理结果决定后续动作．目的地址转换(DNAT, destination network address translation)的回调函数挂载在路由前、输出 Hook 点上，源地址转换(SNAT, source network address translation)的回调函数挂载在输入、路由后 Hook 点上．IPTables 用来维护 NAT 规则．当 1 条新流的数据包到来之后，采取顺序查找的方式匹配 NAT 规则，进行报文头地址转换．随后添加 2 条连接跟踪记录到连接跟踪表中，该流的后续数据包以及反向流的数据包可以直接通过哈希查找的方式查找已有的连接跟踪记录，并根据匹配的记录进行报文头地址转换．

该框架主要有如下缺点：① Netfilter 采用顺序查找算法来匹配 NAT 规则，当规则数目较多时查找匹配时延较大；② NAT 规则和连接跟踪记录均存储在多核共享的全局表中，当多个中央处理器(CPU, central processing unit)核需要同时对 NAT 规则表和连接跟踪表进行查找修改时，读写锁开销制约 NAT 性能的多核扩展性能；③ 网卡接收到数据包后通过中断方式通知内核进行处理，当有大量小包到来时，中断开销很大，极大降低 NAT 的性能．

为了提高 NAT 规则匹配性能，Yang 等[8]采用调整 NAT 规则表中规则顺序的方式，将热点 NAT 规则顺序调到前面，以减少规则查找时间．但是对于非热点 NAT 规则，线性查找开销仍然很大．

为了降低内核处理数据包的开销，多种高性能数据包收发引擎涌现出来．PF-RING[9]实现了内存预分配，为每一个网卡队列分配专一的 CPU 核，减少了内存分配和缓存不一致的开销．PacketShader[10]实现了内存映射，使应用直接访问内核中的数据包缓冲区，减少了内核态到用户态内存拷贝的开销．PFQ[11]采用批处理和并行化技术提升数据包处理性能．数据平面开发套件(DPDK, data plane development kit)[12]采用轮询收包、大页内存、高效资源池、无锁队列等方式，提高了处理大网络流量的效率．然而，这些高性能数据包收发引擎都不包含对数据包包头进行规则匹配并修改的功能，无法直接用于 NAT．

NAT 不仅可以通过软件实现，而且也可以借助专用硬件来实现．Chandrababu 等[13]将规则查找放到专用的网络处理器(NP, network processor)硬件上实现，虽然规则查找性能得到了很大提升，但是增加了部署成本，而且存储规则的数目受到网络处理器硬件容量的限制．专用 NAT 硬件设备可以获得比软件更高的性能[14]，但是硬件价格昂贵且不便于统一管理，而且硬件逻辑比较固定，无法实现灵活多样的业务逻辑[15]．

## 2 通用平台高性能可扩展网络地址转换系统

高性能可扩展网络地址转换系统 Quick NAT 实现在用户态，借助快速 NAT 规则查找算法、并行处理架构、高效数据包处理等方法在通用平台上极大提升了 NAT 性能．

### 2.1 总体架构

Quick NAT 完全在用户态实现，总体架构如图 2 所示，主要由连接跟踪模块、规则查找模块、地址资源池模块、五元组修改模块 4 部分构成．五元组指的是报文头中的源 IP/Port、目的 IP/Port 和传输层协议号．Quick NAT 借助数据平面开发套件 DPDK 网卡轮询驱动进行收包，当新数据包到来后，首先由连接跟踪模块根据报头地址通过哈希查找的方式匹配已有的连接记录，如果找到了相应的连接记录，则说明该数据包属于已有流，后续直接按照连接记录由五元组修改模块对数据包的 IP/Port 地址进行修改，保证同一条流的一致性．如果没有找到对应的连接记录，则说明这个数据包属于新流，于是由规则查找模块通过快速 NAT 规则查找算法(QNS, quick NAT search)来匹配 NAT 规则，并根据对应的 NAT 规则从地址资源池中取出

一个可用的 IP/Port 地址,并添加 2 条连接记录(原方向流和反方向流)来记录该五元组映射关系. 随后五元组修改模块对数据包的 IP/Port 地址进行修改. 当该条流的后续数据包或者反向流的数据包到达后,可以直接根据新添加的连接记录进行五元组修改,而无需再进行 NAT 规则查找,不仅保持了同一条流地址映射的一致性,也提高了网络性能.

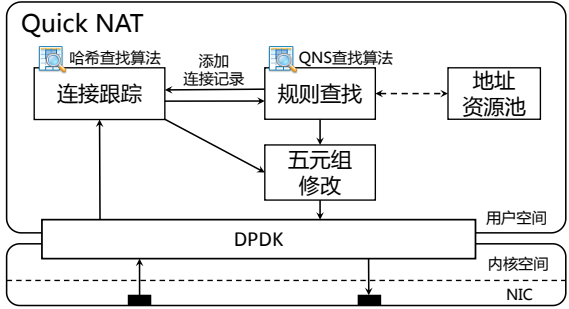

图 2  Quick NAT 总体架构

Quick NAT 的主要贡献为:① 设计了 QNS 算法,基于哈希算法对 NAT 规则进行高速匹配,时间复杂度为 $O(1)$;② 设计了高效并行 NAT 架构,通过多队列网卡接收方缩放 (RSS, receive-side scaling) 和多核 CPU 并行处理多条流,并借助连接记录表本地化和无锁 NAT 规则表来降低多核共享全局连接跟踪表和 NAT 规则表的开销;③ 实现高效数据包处理,采用轮询收包和全程指针操作的方式避免收包中断开销和数据包拷贝开销.

## 2.2 QNS 查表算法

Linux 内核借助 Netfilter/IPTables 架构实现 NAT, 而 Netfilter 通过顺序查找的方式匹配 NAT 规则,复杂度为 $O(n)$,当 NAT 规则较多或网络流量大时开销较大,可扩展性较差. 为了提升性能,基于哈希算法设计了快速 NAT 规则查找算法 QNS,时间复杂度为 $O(1)$,具有很高的可扩展性.

为了更快查找 NAT 规则,需要将 NAT 规则按照 IP 掩码、NAT 规则种类分类存放,因此设计了 32 张 SNAT 子表和 32 张 DNAT 子表来存储 NAT 规则,不同子表对应的 IP 地址掩码和 SNAT/DNAT 种类不同. 为了采用 QNS 算法,需要首先将所有 NAT 规则存入对应的子 NAT 表,图 3 所示为 NAT 规则存储示例. 以图 3(a)第 2 条规则为例,由于其是 SNAT 规则且 IP 掩码为 24,所以需要将其存入掩码 24 的 SNAT 子表中,存储的 Bucket 位置由该规则的三元组(IP、Port、协议号)计算的哈希值来决定,哈希值的范围为[0, 65535],采用链表的方式处理哈希冲突. 当子表中存储有 NAT 规则时,将 Flag 标志位置 1.

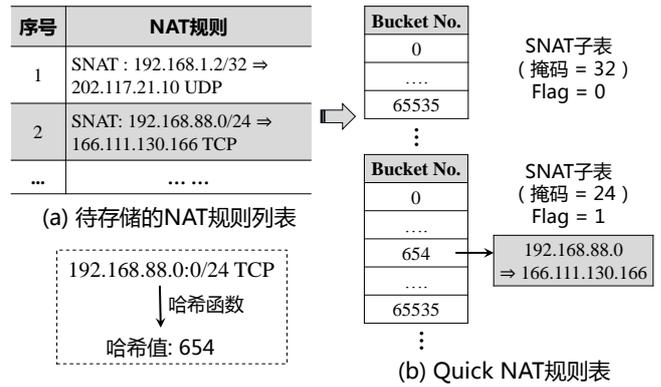

图 3  NAT 规则存储示例

当新数据包到来时,QNS 算法根据数据包的三元组哈希值依次查找各个 DNAT 和 SNAT 子规则表,掩码大的子表匹配优先级更高,在查找子规则表时,首先匹配精确端口的 NAT 规则,随后匹配端口通配的 NAT 规则,当匹配到 NAT 规则后 QNS 算法结束. 按照该算法,如果规则表中有多条匹配的 NAT 规则时,最精确的 NAT 规则先被查找到. 例如,1 个传输控制协议(TCP, transmission control protocol)数据包的源 IP/Port 为 192.168.88.32:1234,目的 IP/Port 为 103.235.46.39:80. 如图 4(a)所示,当查找掩码 32 的 SNAT 子表时,首先对源 IP 地址进行掩码操作,再根据掩码操作后的 IP/Port(192.168.88.32:1234)计算哈希值查找包含源端口的 SNAT 规则,再根据 IP (192.168.88.32)计算哈希值查找端口通配的 SNAT 规则. 在掩码 32 的子表中没有找到匹配规则后,依次查找其他子表. 子表的标志位为 0 表明没有存储 NAT 规则,故跳过该子表以提高查表效率. 当查到掩码 24 的子表时,根据掩码操作后的 IP 地址(192.168.88.0)哈希查找到一条 SNAT 规则,如图 4(b)所示.

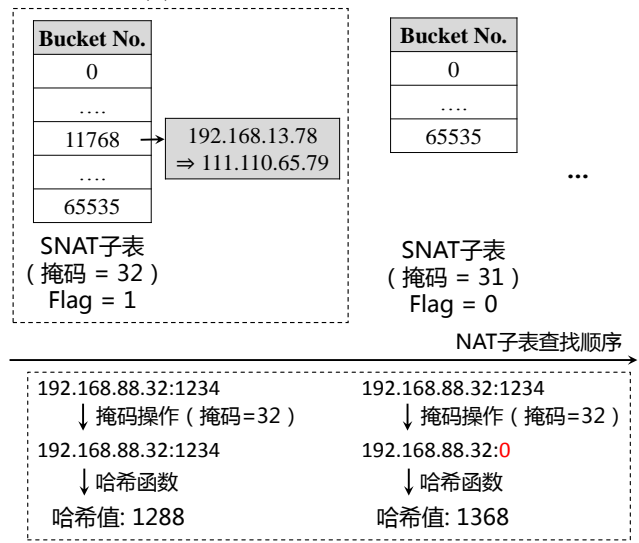

(a) 查找掩码 32 的 SNAT 子表

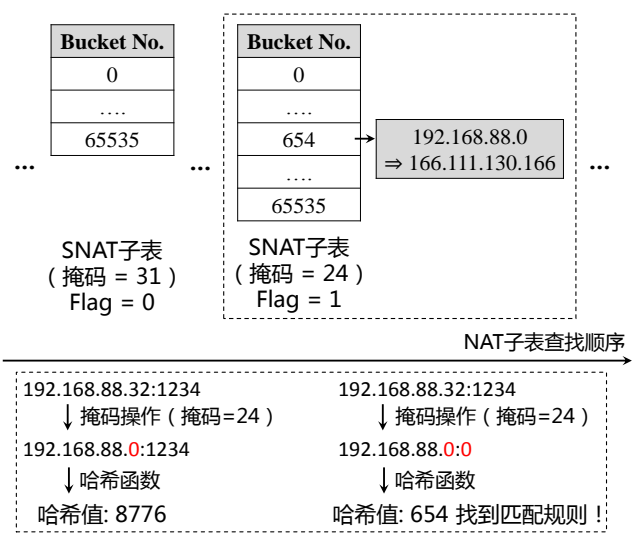

(b) 查找掩码 24 的 SNAT 子表

图 4　NAT 规则匹配示例

### 2.3 高效并行 NAT 架构

随着通用服务器的不断更新换代，多核 CPU 和多队列网卡已经得到了广泛应用，为了提升 NAT 系统在通用平台的性能，需要设计高效并行的 NAT 处理架构，充分利用这些 CPU 和网卡的新特性．

当前内核 Netfilter/IPTables 多核架构借助网卡接收方缩放 RSS 机制将流量均衡到不同 CPU 核并行处理．但是实验结果表明，Netfilter/IPTables 多核性能实际提升有限，主要原因是连接跟踪表和 NAT 规则表是全局共享的，当多个 CPU 核同时对这些表进行查询和修改时，读写锁开销制约着性能的进一步提升，成为多核性能瓶颈．

为了解决这个问题，采用以下方法：

1) 连接跟踪表本地化．将原来全局的连接跟踪表划分为多个本地化的子表，让每个 CPU 核维护管理自己的连接跟踪表，以减少共享全局表带来的锁开销．目前对称 RSS 可以保证双向流均可分配到同一个网卡队列并由同一个 CPU 核进行处理，但是由于 NAT 将流数据包的五元组进行了修改，所以对称 RSS 无法使流量的反向流分配到同一个队列中，即反向流将会由不同的 CPU 进行处理．因此，在根据 NAT 规则做完地址转换后，预先根据网卡 RSS 算法计算出处理反向流的 CPU 核，提前将对应的连接记录添加到该 CPU 核的本地连接跟踪表中．当反向流到达时，对应的 CPU 只需要查询本 CPU 的本地连接跟踪表就可以找到对应的连接跟踪记录，并进行网络地址转换，而无需查询全局连接跟踪表，避免了锁开销．

2) 无锁 NAT 规则表．对于必须全局共享的 NAT 表，采用基于比较并交换原子操作(CAS, compare and swap)的无锁哈希表进行实现，避免锁的开销．

### 2.4 高效数据包处理

内核 Netfilter/IPTables 架构在处理数据包时具有以下开销：① 中断开销．当有新数据包到达网卡时，网卡采用中断的方式通知内核进行处理，当流量较大时，中断开销较大[16]；② 拷贝开销．当内核接收和发送数据包时，需要将数据包在内核与网卡之间进行拷贝，影响网络处理性能的提升．

DPDK 是 Intel 公司开发的数据平面开发工具箱，可以为用户空间高效数据包处理提供函数 API 和驱动支持，具有轮询收包、大页内存、高效资源池、缓存预取、无锁队列、CPU 亲核性绑定等多种优良特性，目前在云计算领域获得了广泛的应用．

为了降低内核 Netfilter/IPTables 架构的中断和拷贝开销，Quick NAT 系统借助 DPDK 的以上特性，在用户态开发部署，采用轮询收包的方式，避免中断收包的高开销．DPDK 驱动程序负责初始化收包描述符，并告知网卡数据包缓冲内存块的物理地址．当有新数据包到达网卡时，网卡会将收到的数据包通过直接存储器访问(DMA, direct memory access)方式拷贝到对应的缓冲内存块中，并将收包标志位置为 1．Quick NAT 系统定期轮询收包标志位即可对收到的数据包进行及时处理．在整个过程中，不会有任何中断产生，而且数据包不需要经过内核拷贝就能直接到达用户态内存空间，在网络流量大时具有较大的性能提升．

针对收到的数据包，Quick NAT 系统通过数据包指针读取数据包的五元组信息，经过 NAT 规则匹配后，如果需要对其五元组进行修改，则直接通过数据包指针修改数据包头相应字段，不需要对数据包进行拷贝，减少了拷贝的开销[17-18]．

在发包时，Quick NAT 系统将包含数据包内存地址的发包描述符放入网卡发包队列中，网卡通过 DMA 方式直接读取内存中的数据包并发送出去．随后 Quick NAT 系统释放已发送数据包所占用的内存空间，并将相应的数据结构放回资源池中，便于高效利用内存资源．

## 3　实验结果

Quick NAT 系统在通用服务器平台上部署和运行，共由大约 4353 行 C 代码组成．服务器系统为 Ubuntu 16.04(kernel 4.4.0)，DPDK 版本为 16.11，CPU 为 Intel Core CPU i7-5930k@3.5 GHz，内存 16 GB（其中分配给大页内存 8 GB），网卡为支持 DPDK 的 Intel 82599ES 10 Gbit/s．一共使用 3 台服务器，1 台部署 Quick NAT 系统，另 2 台分别作为发包和收包设备，收发包程序为风河公司的 DPDK-Pktgen[19]．同时，在 Linux 系统的 Netfilter/IPTables 架构中配置了 NAT 规则，用于对比验证 Quick NAT 在网络地址转换性能方面的优势．

首先对 QNS 查表算法与 Linux 内核 Netfilter/IPTables 架构的线性查表算法性能进行比较．首先存入一定数量的 NAT 规则，随后发送网络流量并记录 NAT 规则查找所需时间．实验结果如表 1 所示，线性查找算法查找时间随着规则数目的增长呈线性增加趋势．而 QNS 随着规则数目的增长查表时间基本不变，而且显著低于线性查找算法，这是由于 QNS 查表复杂度为 $O(1)$，可扩展性较好．

表 1 NAT 规则查表时间　　　　　　　　ns

| 规则条数 | QNS 查找 | 线性查找 |
|---|---|---|
| 100 | 43 | 420 |
| 1000 | 45 | 3256 |
| 3000 | 42 | 9373 |
| 5000 | 45 | 15078 |
| 10000 | 45 | 30120 |

为了对比 Quick NAT 与内核 Netfilter/IPTables 架构的多核可扩展性，在 2 种系统中分别添加了 1000 条 NAT 规则，发送端同时发送 8000 条五元组不同的流，数据包大小为 64 byte，通过改变 NAT 系统使用的 CPU 核数，测量不同数目 CPU 核情况下的吞吐率，结果如图 5 所示．Quick NAT 的吞吐率随着 CPU 核数的增加而不断增长，当 CPU 核数达到 3 个及以上时，吞吐率基本可以达到网卡线速，比内核 Netfilter/IPTables 性能提升 8 倍以上，说明 Quick NAT 具有较强的多核可扩展性．

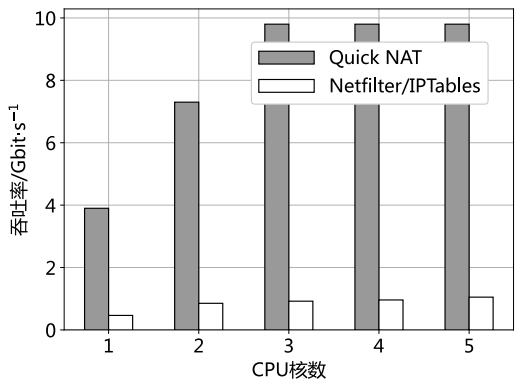

图 5　网络吞吐率随 CPU 核数的变化

为了验证 Quick NAT 在不同数据包长度下的性能提升，在 2 种系统中分别添加了 1000 条 NAT 规则，发送端同时发送 8000 条五元组不同的流，CPU 核数为 5，通过在发送端改变数据包的长度，测量了系统处理不同长度数据包的吞吐率，结果如图 6 所示．内核 Netfilter/IPTables 的吞吐率随着数据包长度的增大而增长，当数据包长度达到 800 byte 及以上时，吞吐率基本可以达到 9.4 Gbit/s．Quick NAT 的吞吐率在不同数据包长度情况下均可达到线速，说明 Quick NAT 具有较强的数据包处理性能．

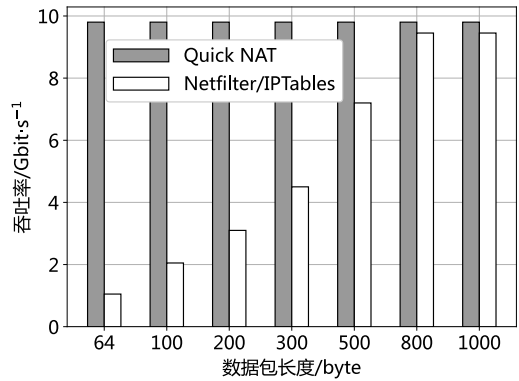

图 6　网络吞吐率随数据包长度的变化

## 4　结束语

为了提高通用平台上 NAT 的性能，设计了高性能可扩展网络地址转换系统 Quick NAT．通过高速 NAT 查表算法 QNS，将原始规则表划分为若干条哈希子表，并采用哈希查表算法极大地提高了规则查找效率；为了充分发挥多核 CPU 与多队列网卡的性能优势，Quick NAT 采用高效并行架构，通过实现连接记录表本地化和基于 CAS 的无锁 NAT 规则表，避免多 CPU 核访问修改全局表带来的锁开销；Quick NAT 部署在用户态，采用轮询取代中断、越过内核等机制，全程使用指针操作数据包实现零拷贝，进一步降低处理开销．实验结果表明，Quick NAT 可以极大地提高 NAT 查表的效率和吞吐量，能够在 10 Gbit/s 网络环境下实现 64 byte 小包线速，比内核 Netfilter/IPTables 性能提升 8 倍以上，具有较强的多核可扩展性．